\documentclass[prb,12pt]{revtex4}

\usepackage[english]{babel}
\usepackage[intlimits]{amsmath}
\usepackage[dvips]{graphicx}

\begin{document}

\begin{titlepage}

\title{Polymer Translocation Through a Long Nanopore}
\author{Elena Slonkina}
\affiliation{Department of Chemistry, Moscow State University, Moscow, Russia 119899}
\author{ Anatoly B. Kolomeisky}
\affiliation{Department of Chemistry, Rice University, Houston, TX 77005}

\begin{abstract} 
Polymer translocation through a nanopore in a membrane investigated theoretically. Recent experiments on voltage-driven DNA and RNA translocations through a nanopore indicate  that the size and geometry of the  pore  are important factors in  polymer dynamics. A theoretical approach is presented  which explicitly takes into account   the effect of the nanopore length and diameter for  polymer motion across the membrane. It is shown that the length of the pore is crucial for polymer translocation dynamics. The present  model predicts that for realistic  conditions (long nanopores and large external fields) there are two regimes of translocation depending on  polymer size: for polymer chains larger than the pore length, the velocity of translocation is nearly constant, while for polymer chains smaller than the pore length the velocity  increases with decreasing polymer size.  These results agree with  experimental data.

\end{abstract}

\maketitle

\end{titlepage}

\section{Introduction}

Translocation of polymers across a nanopore plays a critical role in numerous natural phenomena and industrial processes. Many biological phenomena, such as the motion of DNA and RNA molecules across nuclear pores, virus infection of cells, DNA packaging into viral capsids, gene swapping and protein transport through membrane channels, involve the motion of biopolymers across membranes.\cite{lodish,alberts} In chemistry, the forced permeation of polymer molecules and electrophoresis are crucial for separations and purifications of synthetic as well as biological macromolecules. The motion of polymers in a confined medium is also technologically important in food and medicine production, in oil recovery and separation, and in many other industrial processes. Accordingly, the mechanisms of polymer translocation have become a subject of numerous experimental\cite{KBBD,akeson,meller,henrickson,meller1,movileanu,howorka} and theoretical studies.\cite{sung,dimarzio,degennes,lubensky,muth,muth1,sebastian,konkoli}

A polymer molecule moving across a nanopore faces a large entropic barrier due to the decrease in the number of available configurations for polymer segments. In order to overcome this barrier and to speed up the motion of polymers, an external field or interaction is needed. In recent {\it in vitro} experiments,\cite{KBBD,akeson,meller,henrickson,meller1,movileanu,howorka} DNA and RNA molecules are driven through an $\alpha$-hemolysin membrane channel with the help of an external electric field. These elegant experiments are based on the following  simple idea. When a polymer molecule moves through a nanopore, the electric current in the system nearly vanishes because the polymer blocks the flow of free ions through the channel. Accurate recordings of current blockages allow the description of the dynamics of translocation of {\it single} polymer molecules. The principal experimental findings can be summarized as follows: (i) the ability of polymers to enter the nanopore depends linearly on polymer concentration and exponentially on applied voltage;\cite{KBBD,henrickson} (ii) there is a critical value of the external electric potential below which no polymer molecule can enter and move through the nanopore;\cite{henrickson,meller1} (iii) the effective number of free charges on a translocating polymer is surprisingly very small in comparison with the number of available charges;\cite{henrickson} (iv) there are two regimes of polymer threading through the nanopore depending on polymer length - long polymers move across the membrane with nearly constant velocity, while short polymers move significantly faster;\cite{meller1} (v) the nanopore length defines the boundary between short and long polymers.\cite{meller1} These last two experimental observations are the subject of the present theoretical investigations.

Several theoretical models\cite{sung,dimarzio,degennes,lubensky,muth,muth1,sebastian,konkoli} have been developed in order to explain these experimental findings, however, with limited success. Theoretical approaches to polymer translocation mainly follows three  directions. In one approach,\cite{sung,muth,konkoli} the moving polymer molecule should overcome the entropic barrier, and the free energies of polymer segments determine the dynamics of translocation. Another approach\cite{lubensky}  focuses on the interaction between the polymer and the nanopore, and  neglects the entropic contributions from polymer segments outside  the nanopore.  The last  approach\cite{sebastian} views the polymer translocation as the motion of a kink, which travels in the direction opposite to polymer transport. All these theoretical works provide a reasonable description of  polymer threading through the nanopore for very large polymers. However, these theories  are less successful  in understanding  the dynamics of relatively short polymers due to the fact that they view the nanopore as an object which can hold only one monomer (with the exception of Ref.\cite{konkoli}, as  discussed below). The $\alpha$-hemolysin membrane channel, that has been used as the nanopore in {\it in vitro}  experiments,\cite{KBBD,akeson,meller,henrickson,meller1,movileanu,howorka} has a length of approximately 5 nm for the narrow part of the channel  and thus can hold up to 10-15 DNA or RNA monomers.  The theoretical approach of Ambj\"{o}rnsson et al.\cite{konkoli} takes into account the nanopore length and studies both the polymer entrance into the pore and the translocation process. However, only ideal flexible polymers are considered and theoretical analysis  focused on the dependence of polymer translocation dynamics on external electric field.

In this work  the effects of the nanopore length and diameter on the threading dynamics of single polymer molecules are investigated. The goal is to develop the simplest theoretical description of translocation process which takes into account the geometry of the nanopore and interactions between the nanopore and polymer molecule. In the present model the polymer moves across the membrane as shown in Fig.1. The article is organized as follows. In Section II we develop a model and calculate free energies and translocation times for translocating polymers of different sizes. In Section III we apply our results for the description of experimental translocations of voltage-driven DNA molecules. Our theoretical analysis is summarized in Section IV.

\section{Model for Polymer Translocation}

Consider a polymer molecule consisting of $N$ monomers (each of size $a$), which moves from an upper chamber to a lower chamber through a nanopore of length $l=Ma$ and diameter $d=Da$, as shown in Fig.1. Here we  assume that as soon as the polymer enters into the pore, it is unlikely to come back. This assumption is justified since under  experimental conditions the energy gained by a single monomer by moving through the nanopore is much larger than the thermal energy, and thus the probability to return is very  small.\cite{konkoli} It is  also assumed that the nanopore is  part of an infinite two-dimensional membrane, and  there are no interactions between the polymer and the membrane. Let the chemical potential of the monomer  in the upper region, in the nanopore, and in the lower region be $\mu_{1}$, $\mu_{2}$ and  $\mu_{3}$ respectively (see Fig.1). The potential energy change  is considered to occur  only across the nanopore. 

The simple visual analysis of polymer transport across the nanopore indicates that the motion of long polymers (larger than the nanopore length) is qualitatively different from that  of short polymers, and  these two cases must be considered separately. In our analysis, we assume that both $N$ and $M$ are large, which is consistent with current experimental conditions.\cite{KBBD,akeson,meller,henrickson,meller1,movileanu,howorka} In our theoretical model the  translocation process starts as soon as the first monomer enters the pore, and ends when the last monomer leaves the pore. Note that experimental translocation times are slightly different, as  discussed below.

\subsection{Polymers with sizes $N>M$}

In this case,  there are three regimes of motion, as shown in \mbox {Fig.2a}.  In  regime I, the leading monomer enters the nanopore from the upper region  and then  moves across  the nanopore. In  regime II,  the leading monomer leaves the nanopore, while the end  monomer approaches the entrance of the pore. In regime III, the end  monomer goes through the nanopore and finally leaves it for the lower region. Assuming that the polymer segments inside the nanopore do not contribute to the free energy, i.e., there are no fluctuations inside the pore, the free energy $F_{m}$ of the polymer configuration in regime I with $m$ monomers in the pore and $(N-m)$ monomers in the upper region,  is given by\cite{muth}
\begin{equation}\label{free.energy1}
\frac{F_{m}}{k_{B}T}=(1-\gamma_{1}') \ln (N-m) +\frac{ m \Delta \mu_{1}}{k_{B}T},
\end{equation}
where $\gamma_{1}'$ is a  parameter  which describes the properties of polymers  and which is  equal to 0.5, 0.69 and 1 for Gaussian, self-avoiding, and rod-like chains, respectively;\cite{eisenriegler} the subindex 1 indicates the properties of the upper region of the system. The first term in (\ref{free.energy1}) is an entropic contribution due to $(N-m)$ free monomers in the upper region, while the second term represents the  energy gain due to moving $m$ monomers  into the pore, and  includes the effect of the external field and chemical potential changes. The entropic contribution term follows from  the partition function for the polymer chain in a semi-infinite space near a hard wall with the end monomer anchored at the wall.\cite{eisenriegler,muth}  The chemical potential difference per monomer is given by  $\Delta \mu_{1}=\mu_{2}-\mu_{1}$, and we assume that the potential energy inside the nanopore is uniform. Note that the number of  monomers in the pore $m$ can vary between 0 and $M$. Similarly, the translocation of the polymer in regimes II and III can be described by
\begin{equation}\label{free.energy2}
\frac{F_{m}}{k_{B}T}=(1-\gamma_{1}') \ln (N-M-m)+(1-\gamma_{2}') \ln m +\frac{ m \Delta \mu_{3}}{k_{B}T},
\end{equation}
with $0<m<N-M$, and
\begin{equation}\label{free.energy3}
\frac{F_{m}}{k_{B}T}=(1-\gamma_{2}') \ln m +\frac{ m \Delta \mu_{2}}{k_{B}T},
\end{equation}
with $N-M<m<N$, respectively. Here $\gamma_{2}'$ describes the properties of the polymers  in the lower region, and the chemical potential differences are $\Delta \mu_{2}=\mu_{3}-\mu_{2}$ and $\Delta \mu_{3}=\mu_{2}-\mu_{1}+\mu_{3}-\mu_{2}=\Delta \mu_{1}+\Delta \mu_{2}$.

The transport of the polymer across the nanopore can be described by a Master equation\cite{muth}
\begin{equation}\label{master}
\frac{\partial P_{m}(i,t)}{\partial t}=u_{m-1} P_{m-1}(i,t)+w_{m+1}P_{m+1}(i,t)- (u_{m}+w_{m}) P_{m}(i,t),
\end{equation}
where $P_{m}(i,t)$ is the probability of moving $m$ monomers in regime $i=$I, II or III at time $t$. $u_{m}$ is the rate constant of adding one more monomer to the segment of $m$ monomers already moved, and $w_{m}$ is the rate constant of removing one monomer from the segment of length $m$. These rate constants are related by detailed balance, namely,
\begin{equation}
 \ln \frac{u_{m}}{w_{m+1}}=-\frac{(F_{m+1}-F_{m})}{k_{B}T}.
\end{equation}
Following Muthukumar,\cite{muth} it is assumed  that these rate constants  are independent of $m$; however, they are different for different regimes, i.e., $u_{m}=u_{i}$ for $i=$I, II or III, and generally $u_{1} \neq u_{2} \neq u_{3}$. Transforming the discrete Eq. (\ref{master}) into continuum Smoluchovskii equation, we obtain
\begin{equation}\label{smoluchovskii}
\frac{\partial P_{m}(i,t)}{\partial t}=\frac{\partial}{\partial m} \left[ \frac{u_{i}}{k_{B}T}\frac{\partial F_{m}}{\partial m}  P_{m}(i,t)+ u_{i}\frac{\partial}{\partial m} P_{m}(i,t) \right].
\end{equation}
The mean translocation time $\tau$ can now be calculated as a sum of mean first-passage times in each regime,\cite{risken} i.e., $\tau=\tau_{1}+\tau_{2}+\tau_{3}$, with
\begin{equation}
 \tau_{1}=\frac{1}{u_{1}}\int_{0}^{M}\exp\left(\frac{F_{m_{1}}}{k_{B}T}\right) dm_{1}\int_{0}^{m_{1}}\exp\left(-\frac{F_{m_{2}}}{k_{B}T}\right) dm_{2},
\end{equation}
\begin{equation}
 \tau_{2}=\frac{1}{u_{2}}\int_{0}^{N-M}\exp\left(\frac{F_{m_{1}}}{k_{B}T}\right) dm_{1}\int_{0}^{m_{1}}\exp\left(-\frac{F_{m_{2}}}{k_{B}T}\right) dm_{2},
\end{equation}
\begin{equation}
 \tau_{3}=\frac{1}{u_{1}}\int_{N-M}^{N}\exp\left(\frac{F_{m_{1}}}{k_{B}T}\right) dm_{1}\int_{0}^{m_{1}}\exp\left(-\frac{F_{m_{2}}}{k_{B}T}\right) dm_{2},
\end{equation}
where the corresponding expressions for free energies in different regimes are used. These equations can be solved numerically  for any set of parameters; however, explicit analytic results can be obtained in some special cases. Chemical potential differences are the leading factors in the dynamics of translocation.\cite{muth,muth1} Then,  for $\Delta \mu_{1}= \Delta \mu_{2}= \Delta \mu_{3}=0$, we can obtain exact  expressions for translocation times; namely, $\tau_{i}$ are given by
\begin{equation}
\tau_{1}= \frac{N^{2}}{u_{1} \gamma_{1}'} \left[\frac{1-(1-M/N)^{2-\gamma_{1}'}}{2-\gamma_{1}'}-\frac{1-(1-M/N)^{2}}{2}  \right],
\end{equation}
\begin{equation}
\tau_{2}= \alpha \frac{(N-M)^{2}}{u_{2}},
\end{equation}
\begin{equation}
\tau_{3}= \frac{N^{2}}{u_{3}(2- \gamma_{2}')} \left[\frac{1-(1-M/N)^{\gamma_{2}'}}{\gamma_{2}'}-\frac{1-(1-M/N)^{2}}{2}  \right],
\end{equation}
where $\alpha$ is a constant, which is equal to 1/2 and $\pi^{2}/16$  for the special cases $\gamma_{1}'=\gamma_{2}'=1$ and $\gamma_{1}'=\gamma_{2}'=1/2$, respectively. In the limit  $N \gg M$, these results reduce to
\begin{equation}
\tau_{1} \simeq  \frac{M^{2}}{2 u_{1}}, \quad \quad \tau_{2} \simeq  \frac{\alpha N^{2}}{ u_{2}}, \quad  \quad \tau_{3} \simeq  \frac{M^{2}}{2 u_{3}}. 
\end{equation}
Thus the overall translocation time $\tau$ in this limit is  proportional to $N^{2}$, in agreement with the corresponding results   from Ref.\cite{muth}. For  another limiting case, $ N \sim M$, we obtain in a similar way
\begin{equation}
\tau_{1} \simeq  \frac{N^{2}}{2 u_{1}(2-\gamma_{1}')}, \quad \quad \tau_{2} \simeq  0, \quad  \quad \tau_{3} \simeq  \frac{N^{2}}{2u_{3}\gamma_{2}'}. 
\end{equation}
In this case, the overall translocation time is also proportional to $N^{2}$, however, with a different coefficient.

For more realistic situations, when the  chemical potential differences are  negative and the entropic terms in Eqs.(\ref{free.energy1},\ref{free.energy2},\ref{free.energy3}) are weak in comparison with the $\Delta \mu _{i}$ terms, we obtain in regime I
\begin{equation} \label{tau1.large}
\tau_{1} \simeq \left \{ \begin{array}{cc}
  \frac{k_{B}TM}{u_{1} |\Delta \mu_{1}|}, & M|\Delta \mu_{1}| >1, \\
   \frac{M^{2}}{2 u_{1}}, &  M|\Delta \mu_{1}| < 1, 
  \end{array}  \right.
\end{equation}
in regime II,
\begin{equation} \label{tau2.large}
\tau_{2} \simeq \left \{ \begin{array}{cc}
  \frac{k_{B}T(N-M)}{u_{2} |\Delta \mu_{3}|}, & (N-M)|\Delta \mu_{3}| >1, \\
   \frac{(N-M)^{2}}{2 u_{2}}, &  (N-M)|\Delta \mu_{3}| < 1, 
  \end{array}  \right.
\end{equation}
and in regime III,
\begin{equation} \label{tau3.large}
\tau_{3} \simeq \left \{ \begin{array}{cc}
  \frac{k_{B}TM}{u_{3} |\Delta \mu_{2}|}, & M|\Delta \mu_{2}| >1, \\
   \frac{M^{2}}{2 u_{3}}, &  M|\Delta \mu_{2}| < 1. 
  \end{array}  \right.
\end{equation}
For large positive chemical potential differences we can easily calculate for different regimes
\begin{equation}
\tau_{1} \simeq \frac{1}{u_{1}} \left(\frac{k_{B}T}{u_{1} \Delta \mu_{1}}\right)^{2} \exp\left(M \frac{\Delta \mu_{1}}{k_{B}T}\right),
\end{equation}
\begin{equation}
\tau_{2} \simeq \frac{1}{u_{2}} \left(\frac{k_{B}T}{u_{2} \Delta \mu_{3}}\right)^{2} \exp\left((N-M) \frac{\Delta \mu_{3}}{k_{B}T}\right),
\end{equation}
\begin{equation}
\tau_{3} \simeq \frac{1}{u_{3}} \left(\frac{k_{B}T}{u_{3} \Delta \mu_{2}}\right)^{2} \exp\left(M \frac{\Delta \mu_{2}}{k_{B}T}\right).
\end{equation}
When $N \gg M$, the translocation time $\tau$ is governed by the dynamics in regime II, and it  becomes proportional to the polymer length for large negative chemical potential differences, in agreement with experimental observations.\cite{KBBD,meller1}

\subsection{Polymers with sizes $N<M$}

For relatively short polymers (but recall that $N \gg 1$),  again  three regimes of translocation are observed, as shown in \mbox{Fig. 2}. The motion in  regimes I and III is qualitatively similar to the transport of  long polymers (compare Fig.2a and Fig.2b); however the transport in regime II is {\it different}, since there are no polymer segments in the upper or lower regions. Thus the free energy expressions in regimes I and III are the same as those given by Eqs. (\ref{free.energy1}) and (\ref{free.energy3}), respectively with, however, $0<m<N$ in both regimes. The free energy in regime II can be taken  equal to zero because at this level of approximation we neglect the free energy contribution from the polymer segments  fluctuating inside the nanopore.

Calculations of translocation times can be performed  in a similar fashion as was done for long polymers. First, for translocation time in regime II at all possible values of parameters, it can be  easily computed, 
\begin{equation} \label{tau2.short}
 \tau_{2} = \frac{ (N-M)^{2}}{ 2u_{2}'}, 
\end{equation}
where it is assumed that   $u_{2}' \neq u_{2}$ because   the translocation process is physically  different in this regime for short polymers in comparison with long polymers. 

When $\Delta \mu_{1}=\Delta \mu_{2}=0$, the translocation times in regimes I and III are equal to
\begin{equation}\label{tau1.3.short}
\tau_{1} =   \frac{N^{2}}{2 u_{1}(2-\gamma_{1}')},   \quad  \quad \tau_{3} =  \frac{N^{2}}{2 u_{3}\gamma_{2}'}. 
\end{equation}
Here  we assume that the rate constants  for long and for short polymers are the same, since the dynamics of translocation in these regimes are very similar for short and for long polymers. For  more realistic situations, when $\Delta \mu_{1} <0$ and $\Delta \mu_{2} <0$, and the entropic terms in the  free energy expressions  are small, the calculations in regime I yield
\begin{equation}
\tau_{1} \simeq \left \{ \begin{array}{cc}
  \frac{k_{B}TN}{u_{1} |\Delta \mu_{1}|}, & N|\Delta \mu_{1}| >1, \\
   \frac{N^{2}}{2 u_{1}}, &  N|\Delta \mu_{1}| < 1, 
  \end{array}  \right.
\end{equation}
and in regime III
\begin{equation}
\tau_{3} \simeq \left \{ \begin{array}{cc}
  \frac{k_{B}TN}{u_{3} |\Delta \mu_{2}|}, & N|\Delta \mu_{2}| >1, \\
   \frac{N^{2}}{2 u_{3}}, &  N|\Delta \mu_{2}| < 1. 
  \end{array}  \right.
\end{equation}
For large positive chemical potential differences, translocation times are given by
\begin{equation}
\tau_{1} \simeq \frac{1}{u_{1}} \left(\frac{k_{B}T}{u_{1} \Delta \mu_{1}}\right)^{2} \exp\left(N \frac{ \Delta \mu_{1}}{k_{B}T}\right),
\end{equation}
\begin{equation}
\tau_{3} \simeq \frac{1}{u_{3}} \left(\frac{k_{B}T}{u_{3} \Delta \mu_{2}}\right)^{2} \exp\left(N \frac{\Delta \mu_{2}}{k_{B}T}\right).
\end{equation}

\subsection{Fluctuations inside the nanopore}

So far in our calculations of polymer translocation times we neglected the contributions from the fluctuations of polymer segments inside the nanopore, although these fluctuations  may be important. To take them into account   the scaling analysis can be used to describe   the polymer molecule  inside the  confined cylindrical pore.\cite{degennes1}  The free energy of confined polymer chain is given by $k_{B}T N_{b}$, where $N_{b}=l/d$ is the number of blobs inside the pore.\cite{degennes1} Note, however, that this approach is valid when $d \ll l$. The size of each blob is equal to the diameter of the pore, i.e.,
\begin{equation}
d=Da=ag^{\nu},
\end{equation}
where $g$ is the number of monomers in the blob, and the exponent $\nu$ is equal to 1/2, 3/5, and 1 for ideal, self-avoiding, and rod-like chains, respectively. Then the maximum number of monomers in the pore of length $l=Ma$ and diameter $d=Da$ is given by
\begin{equation}
M_{max}=\frac{l}{d} g=MD^{(1/\nu -1)}.
\end{equation} 
Note that for rod-like chains $M_{max}=M$, while for ideal flexible chains $M_{max}=MD$.

Knowing the free energy contribution of polymer segments inside the nanopore allows to  calculate the translocation dynamics as discussed  in detail above. Consider first the dynamics of the polymer molecule in regime II. The contribution from the fluctuating polymer segments inside the pore is always constant in this regime because the number of monomers inside the nanopore does not change. Then this free energy term will not affect translocation times since they are determined by free energy differences [see Eqs.(\ref{tau1.large}),(\ref{tau2.large}),(\ref{tau3.large})]. In regimes I and III the number of monomers inside the nanopore is changing, however the free energy difference from this confinement term is equal to $k_{B}TD^{-1/\nu}$, which for experimental conditions\cite{KBBD,henrickson,meller} is very small in comparison with entropic and chemical potential terms, and  can be neglected. Thus the free energy contributions from fluctuating monomers inside the nanopore do not change the results  on translocation dynamics of the polymers (provided that $M$ is replaced by $M_{max}$).

\section{Comparison with Experiments}

The above results are well compared with the experimental findings of Ref.\cite{meller1}, where the size dependence of voltage-driven single-stranded DNA molecules has been investigated. In the present theoretical approach, the process of translocation  is assumed to start as soon as the leading monomer enters the nanopore and to end when the end monomer leaves the nanopore. Then the translocation velocity  is given by
\begin{equation} \label{velocity}
V=(N+M)a/ \tau.
\end{equation}
For realistic situations (large $N$ and $M$,  $\Delta \mu_{i} \ll 0$), the results for translocation times in corresponding regimes [see Eqs. (\ref{tau1.large}),(\ref{tau2.large}),(\ref{tau3.large}),(\ref{tau2.short}) and (\ref{tau1.3.short})] can be substituted  into equation (\ref{velocity}), leading to explicit expressions for the translocation velocity
\begin{equation} \label{theoretical.velocity}
V =\left \{ \begin{array}{cc}
  (N+M)a/\left(\frac{k_{B}TM}{u_{1} |\Delta \mu_{1}|}+\frac{k_{B}T(N-M)}{u_{2} |\Delta \mu_{3}|}+\frac{k_{B}TM}{u_{3} |\Delta \mu_{2}|}\right), & N > M, \\
 (N+M)a/\left(\frac{k_{B}TN}{u_{1} |\Delta \mu_{1}|}+\frac{(M-N)^{2}}{2u_{2}'}+\frac{k_{B}TN}{u_{3} |\Delta \mu_{2}|}\right)  , &  N < M. 
  \end{array}  \right.
\end{equation}
However, in the experiments of Meller  et al.,\cite{meller1} the translocation time was measured only  when a  current passing through the nanopore  dropped to a level below 65$\%$ of an open channel current. The authors  also showed that the blockade level is proportional to the fractional volume of the channel occupied by the polymer. This means that in these experiments the translocation process started when 35$\%$ of the polymer  entered into the nanopore, and  ended when only 35$\%$ of the polymer  left in the pore.  Thus, in order to compare our theoretical predictions with experimental observations,  the expressions for translocation velocity (\ref{theoretical.velocity}) should be modified as follows
\begin{equation}\label{velocity.exp}
V =\left \{ \begin{array}{cc}
  (N+0.30M)a/\left(\frac{0.65Mk_{B}T}{u_{1} |\Delta \mu_{1}|}+\frac{k_{B}T(N-M)}{u_{2} |\Delta \mu_{3}|}+\frac{0.65M k_{B}T}{u_{3} |\Delta \mu_{2}|}\right), & N > M, \\
 (N+0.30M)a/\left(\frac{k_{B}T(N-0.35M)}{u_{1} |\Delta \mu_{1}|}+\frac{(M-N)^{2}}{2u_{2}'}+\frac{k_{B}T(N-0.35M)}{u_{3} |\Delta \mu_{2}|}\right)  , &  N < M. 
  \end{array}  \right.
\end{equation}

Under experimental conditions,\cite{meller1} the single-stranded DNA  molecules behave more like rod-like polymers, and this fact justifies using $M_{max}=M$ in our description of experimental data. Then the  expressions (\ref{velocity.exp}) can be used to fit the observed translocation velocities\cite{meller1} as shown in Fig.3. The present  theoretical approach predicts two types of translocation depending on polymer size. For large polymers, larger than the nanopore length, the translocation velocity approaches a constant value, while for short polymers the velocity  increases significantly with decreasing polymer length. These predictions are in excellent qualitative and quantitative agreement with experiments for large polymers; however, for short polymers the agreement is only qualitative. 

There are several reasons to explain the deviations between the presented  theory and experimental behavior for short polymers. In our theoretical approach we used a polymer description of molecule dynamics, while for such short polynucleotides ($N=4$---12) the polymer description is probably less precise and the discrete chemical nature of the molecules should be taken into account. In addition, our descriptions of the nanopore geometry and the potential changes inside the nanopore are very simplified. However,  for  short polymer molecules these factors probably influence the translocation dynamics much stronger than for  large polymers. Despite these discrepancies, the fact that a  very simple theoretical approach can provide a qualitative and semi-quantitative description of complex translocation processes is rather encouraging. It also indicates that the presented theoretical  model correctly captures and  describes the main features  of  translocation phenomena.

Our theoretical approach allows us  to investigate the effect of interactions between the nanopore and the polymer molecule. The rate constants $u_{j}$ measure the degree of such interactions. The smaller the rate constants, the larger the attraction between the moving polymer chains and the nanopore. As shown in Fig.4, when only the rate constants  for short polymers in regime II are varied, the interactions between the nanopore and the polymer can change the translocation dynamics significantly. The stronger the interaction, the slower the motion of threading polymer molecules, in agreement with intuition.

\section{Summary and Conclusions}

A simple theoretical model of polymer translocation through the long nanopores driven by external electric fields is presented. The fact that we take into account the nanopore length and diameter allows us to describe the translocation dynamics for polymer molecules of different sizes. By considering in detail different regimes of moving polymers across the membranes, the general expressions for free energies and translocation times for polymer chains threading through the nanopores are derived. The presented theoretical predictions are compared with  experimental results on voltage-driven translocations of single-stranded DNA molecules through the $\alpha$-hemolysin protein channels.\cite{meller1} It is found that for experimental conditions,\cite{meller1} long  polymers, longer than the nanopore length, translocate with nearly constant velocity, while short polymers move significantly faster.   The theoretical analysis indicates that for experiments\cite{meller1} with  $\alpha$-hemolysin protein channels  the polymer fluctuations inside the nanopore do not affect the translocation dynamics. Presented theoretical  results are in  agreement with experimental observations. 

Although  a reasonable description of polymer translocation experiments\cite{meller1} is provided, there are many factors that  have not been taking into account. We  assumed in our calculations that the external field inside the nanopore is uniform, while more realistic picture would incorporate a potential profile inside the nanopore,\cite{konkoli} which can be found by taking into account the realistic geometry of the nanopore and all electrostatic interactions. In the present  model, the possibility that the polymer molecule can return was neglected, which is a very good approximation at large external driving fields, as realized in most experiments. Given theoretical  approach allows  to consider this effect by solving the Smoluchovskii equations (\ref{smoluchovskii}) with different boundary conditions. We also assumed that the nanopore is very narrow, i.e., the effect of the nanopore diameter on free energies of the polymer segments outside of the pore has not been considered, although the  $\alpha$-hemolysin pore in principle  can hold several monomers of DNA or RNA molecules. Probably, the simplest way to  include this possibility into presented theoretical model, is to utilize the  scaling approach.\cite{degennes1}

\section*{Acknowledgments}

ABK is grateful to Dr. A. Meller for introducing to this  problem and for valuable discussions. The critical comments by  Profs. M. Robert,  M.E. Fisher, and  M. Pasquali, Dr. P. Willis and A. Montesi  are  highly  appreciated. The financial support of the Camille and Henry Dreyfus New Faculty Awards Program (under Grant NF-00-056) is  gratefully acknowledged. The authors are also  grateful to Center of Biological and Environmental Nanotechnology at Rice University for financial support.

\newpage
\section*{\bf Figure Captions}

\noindent \mbox{Fig. 1}. \hspace{0.4em} A polymer molecule moves from the upper chamber to lower through a cylindrical nanopore of length $l$ and diameter $d$. The potential energy change is considered to be only inside the nanopore. In polymer translocation experiments (see Refs. \cite{KBBD,akeson,meller,henrickson,meller1,movileanu,howorka}), $l \simeq 5$ nm, $d \simeq 2$ nm, and  $\Delta V \simeq 50$---300 mV.

\vspace{5mm}

\noindent \mbox{Fig. 2}. \hspace{0.4em} Three regimes of translocation for polymers of different sizes: a) for long polymers, b) for short polymers. The size of the nanopore is $l=Ma$. Solid filled circles indicate the head  and the end monomer of the polymer molecule.

\vspace{5mm}

\noindent \mbox{Fig. 3}. \hspace{0.4em} Translocation velocity as a function of  polymer size. The length of a nanopore is equal to 12$a$, where the monomer size is given by  $a=4 \times 10^{-4} \ \mu m$. Filled squares are experimental observations from Ref. \cite{meller1} obtained at external field of 120 mV. Solid lines are our fits with $\Delta \mu_{1}=\Delta \mu_{2}=2.5 k_{B}T$, $u_{1}=u_{3}=2.3 \times 10^{4}$ s$^{-1}$, $u_{2}=1.4 \times 10^{4}$ s$^{-1}$, and $u_{2}'=4.5 \times 10^{6}$ s$^{-1}$. Eq. (\ref{velocity.exp}) is used to calculate theoretical curves.

\vspace{5mm}

\noindent \mbox{Fig. 4}. \hspace{0.4em} Translocation velocities for different degrees of interactions between the nanopore and the polymers. Solid curve is the same as in Fig.3 with $u_{2}'=4.5 \times 10^{6}$, while dashed curve is for the case when $u_{2}'=1.0 \times 10^{6}$, dot-dashed curve is for the case when $u_{2}'=0.5 \times 10^{6}$, and dotted curve is for the case when $u_{2}'=0.2 \times 10^{6}$.

\begin{figure}[t]
\begin{center}
\vskip 1.5in
\unitlength 1in
\begin{picture}(4.5,3.3)
\resizebox{4.5in}{3.3in}{\includegraphics{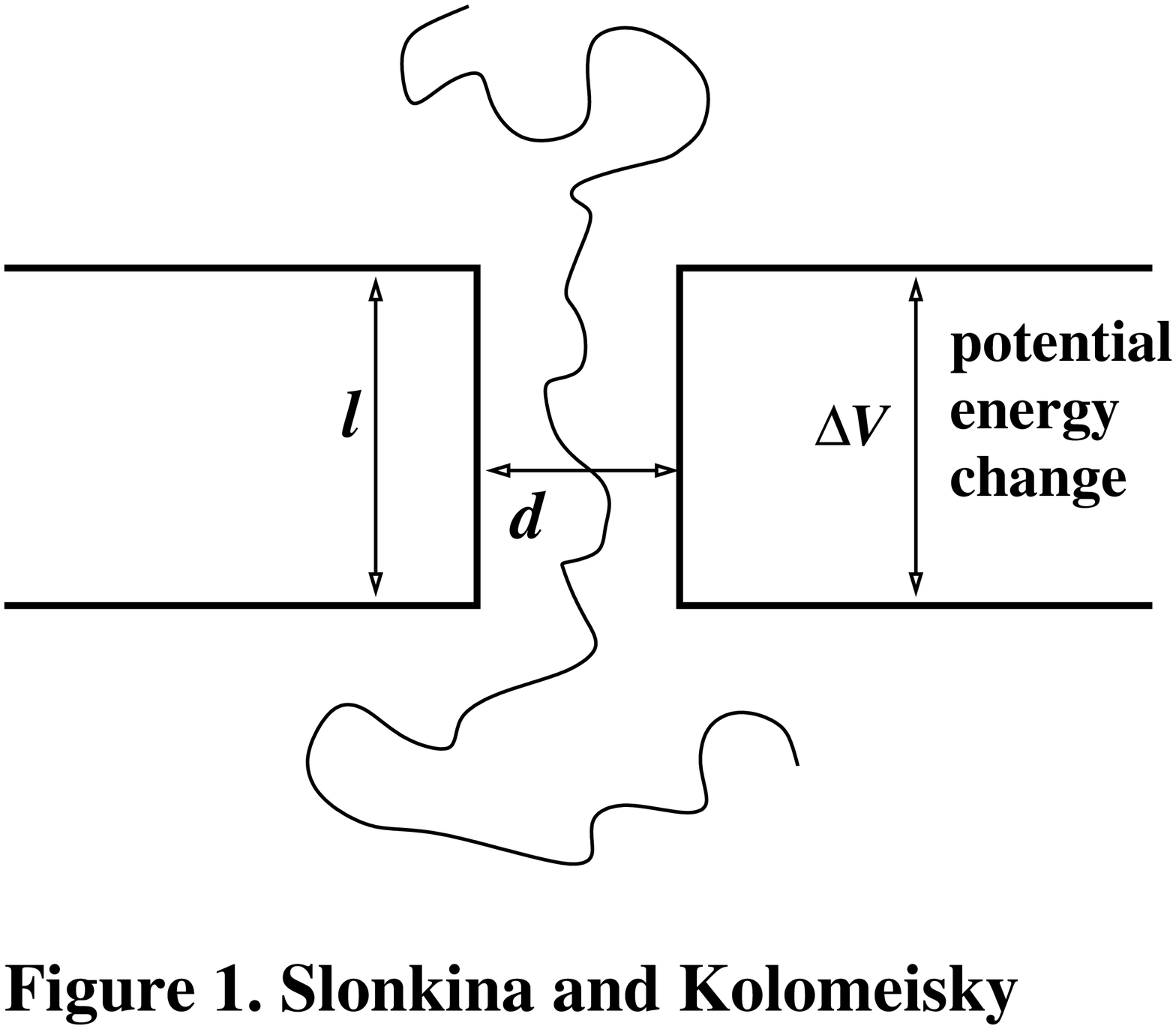}}
\end{picture}

\end{center}

\end{figure}

\begin{figure}[t]
\begin{center}
\vskip 1.5in
\unitlength 1in
\begin{picture}(4.5,3.3)
\resizebox{4.5in}{3.3in}{\includegraphics{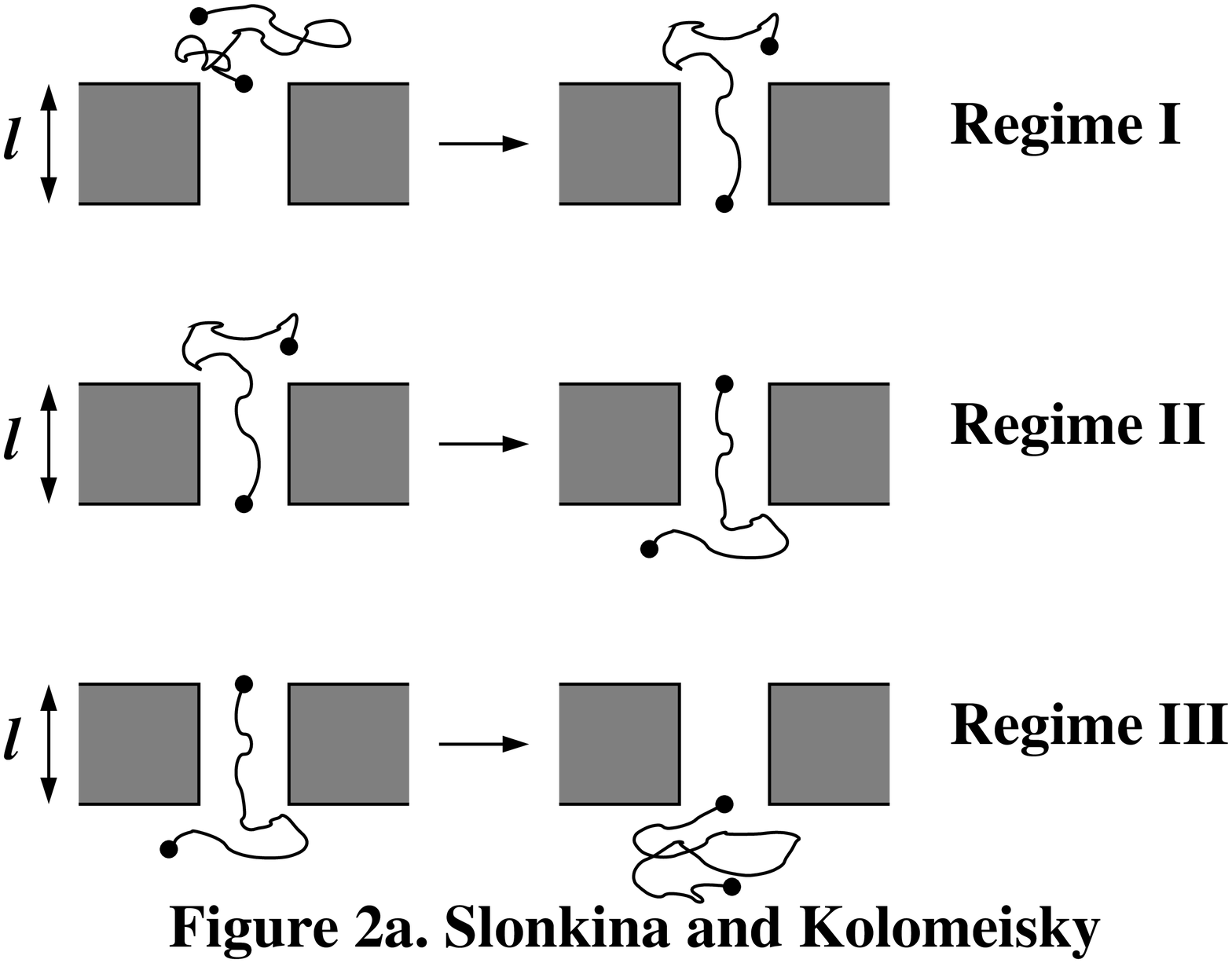}}
\end{picture}

\end{center}

\end{figure}

\begin{figure}[t]
\begin{center}
\vskip 1.5in
\unitlength 1in
\begin{picture}(4.5,3.3)
\resizebox{4.5in}{3.3in}{\includegraphics{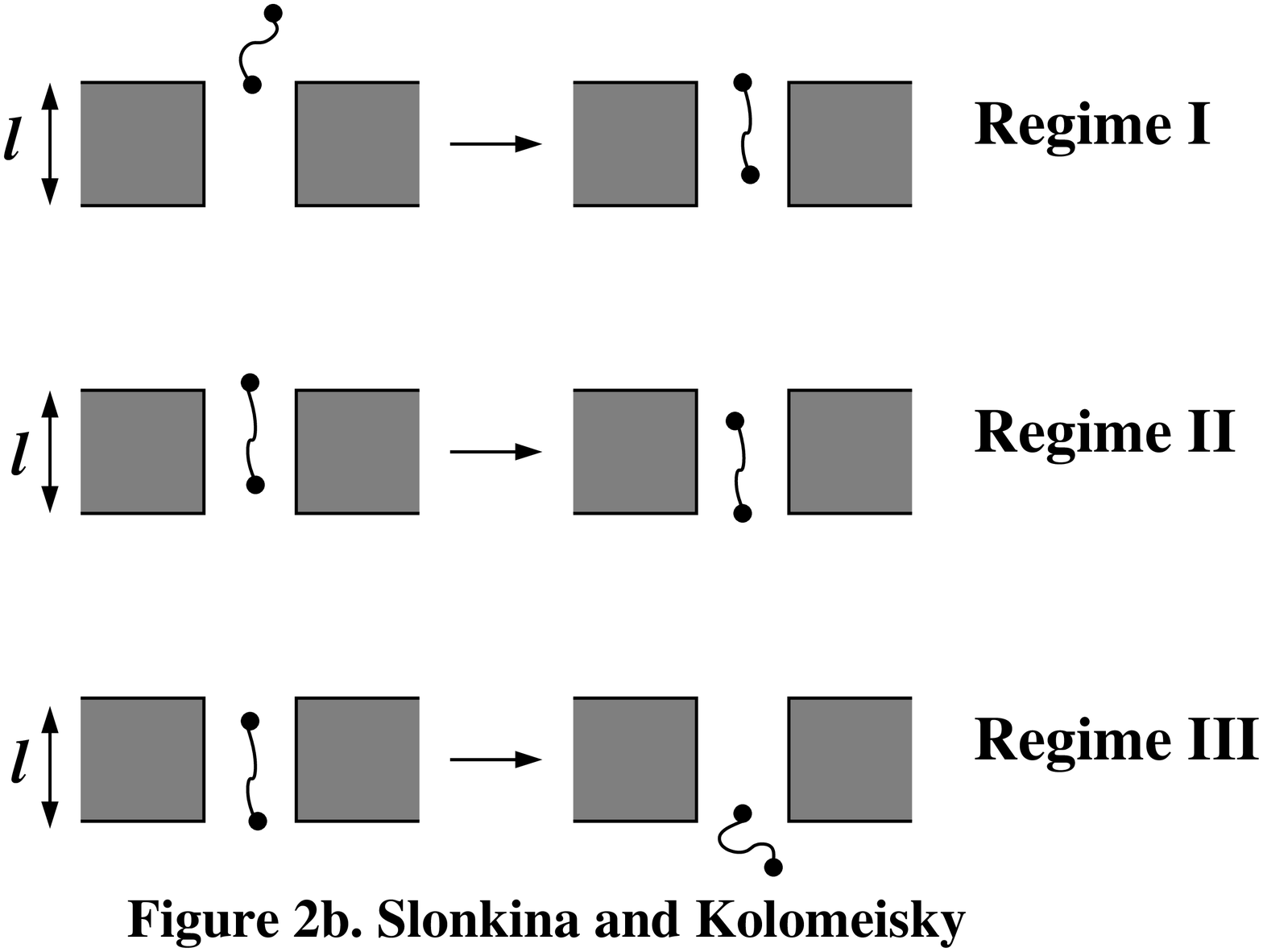}}
\end{picture}

\end{center}

\end{figure}

\begin{figure}[t]
\begin{center}
\vskip 1.5in
\unitlength 1in
\begin{picture}(4.5,3.3)
\resizebox{4.5in}{3.3in}{\includegraphics{Fig3.nanopore.eps}}
\end{picture}

\end{center}

\end{figure}

\begin{figure}[t]
\begin{center}
\vskip 1.5in
\unitlength 1in
\begin{picture}(4.5,3.3)
\resizebox{4.5in}{3.3in}{\includegraphics{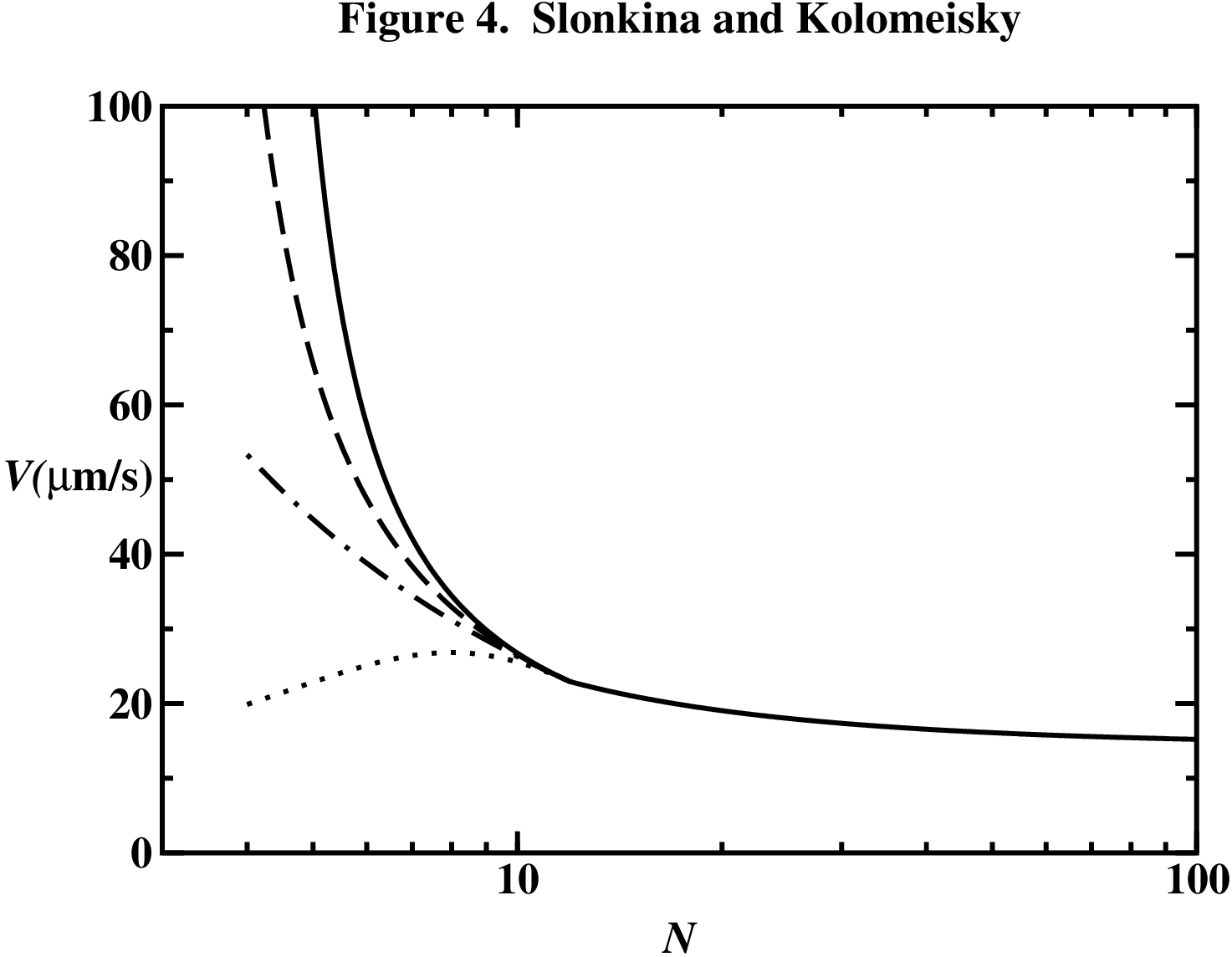}}
\end{picture}

\end{center}
\end{figure}

\end{document}